\documentclass[confernece]{IEEEtran}
\IEEEoverridecommandlockouts
\usepackage[left=0.625in, right=0.625in, top=0.75in, 
bottom=1in]{geometry}
\usepackage{cite}
\usepackage{amsmath,amssymb,amsfonts}
\usepackage{graphicx}
\usepackage{textcomp}
\usepackage{xcolor}
\usepackage{lineno}
\usepackage{changes}
\usepackage[ruled, lined, longend, linesnumbered]{algorithm2e}
\usepackage{amsmath, amsthm, amssymb}

\usepackage{amsthm}
\usepackage{placeins}
\usepackage{csquotes}
\usepackage{url}
\usepackage{multirow}
\usepackage{amssymb}
\usepackage{graphics}
\usepackage{subfigure}
\usepackage{caption}
\usepackage{algpseudocode}
\usepackage{amsfonts}
\usepackage{filecontents}
\usepackage{mathtools}
\usepackage{enumitem}
\def\BibTeX{{\rm B\kern-.05em{\sc i\kern-.025em b}\kern-.08em T\kern-.1667em\lower.7ex\hbox{E}\kern-.125emX}}

\newcommand{\phib}{{\mbox{\boldmath $\phi$}}}

\newcommand{\thetab}{{\mbox{\boldmath $\theta$}}}

\newsavebox\mybox

\begin{document}
\title{SLIPT in Joint Dimming Multi-LED OWC Systems with Rate Splitting Multiple Access}
\vspace{-0.50cm}
\author{\IEEEauthorblockN 
		{Sepideh~Javadi$^{\dag}$, Sajad~Faramarzi$^{\S}$, Farshad~Zeinali$^{\dag}$, Hosein~Zarini$^{\S\S}$, ~Mohammad~Robat~Mili$^{\dag}$,\\~Panagiotis D. Diamantoulakis$^{\star}$,~Eduard Jorswieck$^{\dag\dag}$,~George K. Karagiannidis$^{\star, \star\star}$
		}
		\vspace{0.50cm}
		\\$^{\dag}$ Pasargad Institute for Advanced Innovative Solutions (PIAIS), Tehran, Iran
		\\$^{\S}$ Dept. of Electrical Engineering, Iran University of Science and Technology, Tehran, Iran\\
		$^{\S\S}$ Dept. of Computer Engineering, Sharif University of Technology, Tehran, Iran\\
		$^{\star}$ Dept. of Electrical and Computer Engineering,  Aristotle University of Thessaloniki, Thessaloniki, Greece\\
		$^{\star\star}$ Artificial Intelligence \& Cyber Systems Research Center, Lebanese American University (LAU), Lebanon\\
		$^{\dag\dag}$ Institute for Communications Technology, Technische Universitat Braunschweig, Braunschweig, Germany\\
	}
	\maketitle
\begin{abstract}
Optical wireless communication (OWC) systems with multiple light-emitting diodes (LEDs) have recently been explored to support energy-limited devices via simultaneous lightwave information and power transfer (SLIPT). The energy consumption, however, becomes considerable by increasing the number of incorporated LEDs. This paper proposes a joint dimming (JD) scheme that lowers the consumed power of a SLIPT-enabled OWC system by controlling the number of active LEDs. We further enhance the data rate of this system by utilizing rate splitting multiple access (RSMA). More specifically, we formulate a data rate maximization problem  to optimize the beamforming design, LED selection and RSMA rate adaptation that guarantees the power budget of the OWC transmitter, as well as the quality-of-service (QoS) and an energy harvesting level for users. We propose a dynamic resource allocation solution based on proximal policy optimization (PPO) reinforcement learning. In simulations, the optimal dimming level is determined to initiate a trade-off between the data rate and power consumption. It is also verified that RSMA significantly improves the data rate.   
\end{abstract}
\begin{IEEEkeywords}
Optical wireless communication (OWC), simultaneous lightwave information and power transfer (SLIPT), joint dimming (JD), rate splitting multiple access (RSMA), proximal policy optimization (PPO).\vspace{-0.20cm}
\end{IEEEkeywords}
\IEEEpeerreviewmaketitle
\section{Introduction}
Benefiting from illumination and communication at the same time, optical wireless communication (OWC) systems are envisioned to be a promising technology to compensate for the shortcomings of conventional radio-frequency (RF) communication systems. This technology requires a light-emitting diode (LED) for signal transmission at the transmitter and photo-diode (PD) for signal decoding at the receiver \cite{Surv, Komine, zarini, Pathak}.
So far, advanced optical wireless techniques have been investigated to unleash the potentials of OWC systems, such as multi-LED transmitters \cite{one}, joint dimming (JD) \cite{five}, as well as simultaneously lightwave information and power transfer (SLIPT) \cite{two, Ma, Abdelhady}. In the former case, multiple LEDs are incorporated in an LED array, commonly known as a spatial multiplexing OWC system. This extension brings about remarkable data rate and extended coverage over the OWC systems compared to the single LED. The second case, as a multi-domain control scheme, relies on both analog dimming (AD) and spatial dimming (SD). 
\\Introduced first in \cite{five}, JD control jointly optimizes the direct-current (DC) bias level (in AD), as well as the number of glared LEDs (in SD) to satisfy a required dimming level \cite{six},  \cite{eight}. 
The latter case, i.e., the SLIPT technology, is a promising solution for low-battery devices \cite{two}, which enables OWC receivers to simultaneously obtain illumination, information, and energy harvesting (EH) via a PD, a solar panel, or both. 
\\The concept of wireless information and power transfer has been throughly investigated in the literature \cite{WIPT, J1, J2, J3, J4}. However, the coexistence of alternating current (AC) and DC signals in the photo-current of LEDs has encouraged the researchers to study the SLIPT technology \cite{Liu}. Previously, the performance of an OWC system with multi-LED transmitter and SLIPT technology was considered in \cite{three} and \cite{four} to minimize the energy consumption and maximize the data rate, respectively. These techniques consume significant energy. Furthermore, they were proposed under the assumption of orthogonal resource allocation which limits the data rate. 
In this paper, we study the performance of \cite{three} and \cite{four} by exploring rate splitting multiple access (RSMA)~\cite{PPO2, Bruno1, Bruno2, Bruno3}, which employs non-orthogonality of resources to enable higher data rate and massive access. Additionally, we adopt an efficient JD control scheme, where the number of active LEDs are controlled, thereby reducing the energy consumption of \cite{three} and \cite{four}. 
\\To analyze the performance of this system, a data rate maximization problem is formulated by jointly optimizing the transmit beamforming, LED selection, and RSMA rate adaptation. This problem ensures the power budget of the multi-LED transmitter, as well as receivers' quality-of-service (QoS), dimming level and EH level.
More specifically, we transform the problem into a Markov decision process (MDP) and propose a real-time dynamic solution methodology, based on proximal policy optimization (PPO) reinforcement learning. Through numerical simulations, we compute the optimal dimming level for a trade-off between the data rate and power consumption. Additionally, leveraging RSMA enhances the data rate of this system, compared to the non-orthogonal multiple access (NOMA).
\\The remainder of this paper is organized as follows. The system model of the proposed downlink SLIPT-assited JD multi-LED OWC network with RSMA is introduced in Section II, while Section III evaluates the performance of the proposed DRL-based approach in detail by employing PPO reinforcement learning. In Section IV, the effectiveness of the proposed OWC system is verified by simulation results. Finally, Section V concludes the paper.
\par\emph{Notation:} The vectors and matrices are specified by boldface lower-case and upper-case letters, respectively, while $\text{diag(.)}$ represents the diagonalization operation. The absolute value of scalar $a$, the transpose, and the Hermitian transpose of the vector $\bold{a}$ are denoted by $|a|$, $\bold{a}^T$, and $\bold{a}^H$, respectively. Finally, the expectation operator and the set of real numbers are denoted by $\mathbb{E} (.)$ and $\mathbb{R}$, respectively,  while $[.]$ represents the round off operation.    
\section{System Model}
As illustrated in Fig. 1, we consider the downlink transmission of an OWC network, in which the transmitter is equipped with an LED array, including a set $\mathcal{N} = \left\{ {1,2,...,{N}} \right\}$ of ${N}$ LEDs and communicates with a set $\mathcal{K} = \left\{ {1,2,...,{K}} \right\}$ of $K$ single-PD users. The users are supposed to be distributed randomly, whereas the LEDs are independently modulated via separate drivers, yet all are connected to a central controller that collects channel feedback and performs resource management. 
\begin{figure}
	\centering
	\includegraphics[width=8cm, height=8cm]{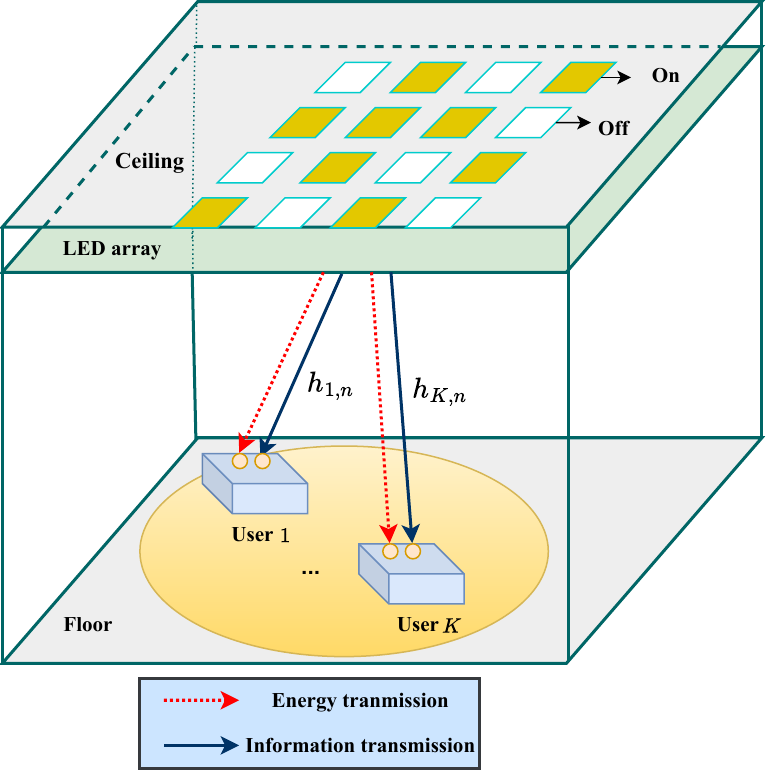}
	\caption{\footnotesize{An OWC system with an LED array, serving $K$ single-PD users for illumination, communication and EH.}}
	\label{Fig5}
\end{figure}
\vspace{-0.3cm} 
\subsection{Signal Model} 
We adopt RSMA as the state-of-the-art multiple access scheme. On this basis, the transmit lightweight data stream for each user has a two-fold structure, including a common message in addition to a private message. The former, i.e., the common message of the transmit lightweight data stream has the same content for all users, whereas the latter, i.e., the private message is exclusively encoded for each user. In other words, one common lightweight transmit data stream, as well as $K$ private ones, form a superimposed transmit signal, carrying $K+1$ messages.
Let ${s}^{\textrm{(c)}}$ and ${{s}}_{k}^{\textrm{(p)}}, \forall k \in \mathcal{K}$, denote the common message shared among all users and the private message of the $k$-th user, respectively, such that $\mathbb{E}\lbrace \lvert {s}^{\textrm{(c)}} \rvert^{2} \rbrace = 1$ and $\mathbb{E}\lbrace \lvert {{s}}_{k}^{\textrm{(p)}} \rvert^{2} \rbrace =
1, \forall k \in \mathcal{K}$. We devise a linear precoding scheme \cite{one} prior to signal transmission, to handle the inter-user interference. Next, a DC bias ${\textbf{i}_{\textrm{DC}}} =
[{i_{\textrm{DC}}},...,{i_{\textrm{DC}}}]^{N \times 1}$ is added to the precoded signal before transmission. This bias regulates the brightness of the LEDs and guarantees that the amplitude of the transmitted signal has a real non-negative value. Accordingly, the lightweight transmit data stream of all LEDs, denoted by $\textbf{x} = {\left[ {{x_1},{x_2},...,{x_N}} \right]^T}$, can be expressed as
\begin{align}
\textbf{x}=
{\textbf{w}}^{\textrm{(c)}}{s}^{\textrm{(c)}}+\sum_{k=1}^{K}{\textbf{w}}^{\textrm{(p)}}_{k}{{s}}_{k}^{
\textrm{(p)}}+\textbf{i}_{\text{DC}},
\end{align}
where ${\bold{w}^{\textrm{(c)}}} = {\left[ {{w_{1}^{\textrm{(c)}}},{w_{2}^{\textrm{(c)}}},...,{w_{N}^{\textrm{(c)}}}} \right]^T} \in {\mathbb{R}^{{N} \times 1}}$ and ${\bold{w}_k^{\textrm{(p)}}} = {\left[ {{w_{k,1}^{\textrm{(p)}}},{w_{k,2}^{\textrm{(p)}}},...,{w_{k,{N}}^{\textrm{(p)}}}} \right]^T} \in {\mathbb{R}^{{N} \times 1}}, ~\forall k \in\mathcal{K}$ specify the common transmit beamforming for ${s}^{\textrm{(c)}}$ and the $k$-th private transmit beamforming for ${{s}}_{k}^{\textrm{(p)}}$, respectively.
Note that ${\textbf{i}_{\text{DC}}}$ has the same value for all LEDs, because of uniformity of illumination in indoor environments \cite{one}. The dynamic range of the LEDs is constraint to avoid signal clipping \cite{ten}. In other words,
\begin{align}\label{abcd}
\left| {{w_{n}^{\textrm{(c)}}}} \right| + \sum\limits_{k = 1}^K {\left|
{{w_{k,n}^{\textrm{(p)}}}} \right|} \le \Xi, ~\forall n \in
\mathcal{N},
\end{align}
where $\Xi=\min ({i_{\text{DC}}} - {I_l},{I_h} - {i_{\text{DC}}})$, and the notations ${I_l}$ and ${I_h}$ indicate the minimum and maximum permissible
currents of all LEDs, respectively.
\subsection{Channel Model}
In this paper, we only consider the line of sight (LoS) links \cite{five}-\cite{eight}. The optical channel gain between the $n$-th LED and the $k$-th user, denoted by $h_{k,n}\in \mathbb{R}$, can be modelled as
\begin{equation}
		h_{k,n}\!=\!\left\{
		\begin{array}
		[c]{c}%
		\!\!\!\!\frac{(m+1)A^{\textrm{OWC}}}{2\pi d_{k,n}^{2}}G^{\textrm{OWC}}(\psi_{k,n})Z^{\text{OWC}},\\
		0,
		\end{array}
		\right.  \left.
		\begin{array}
		[c]{c}%
		\!\!0\!\leq\!\psi_{k,n}\!\leq\!\Psi_{c},\\
		\psi_{k,n}>\Psi_{c},%
		\end{array}
		\right.
\end{equation}
where $Z^{\text{OWC}}=\!\cos^{m}(\phi_{k,n})\!\cos(\psi_{k,n})$, while $A^{\textrm{OWC}}$ and $d_{k,n}$ denote the physical area of the PD for each user and the distance between the $n$-th LED and the $k$-th user, respectively. Moreover, $m=-\frac{\ln2}{\ln(\cos
	\Phi_{1/2})}$ specifies the Lambertian emission order with $\Phi_{1/2}$ being the semi-angle at half-power of the LED. Besides, the angles of incidence and irradiance are respectively given by $\psi_{k,n}$ and $\phi_{k,n}$, while the receiver field of vision (FOV) semi-angle is denoted by $\Psi_{c}$. Finally, $G^{\textrm{OWC}}(\psi_{k,n})$ indicates the gain of the optical concentrator which is defined as follows
	\begin{equation}
	G^{\textrm{OWC}}(\psi_{k,n})=\left\{
	\begin{array}
	[c]{c}%
	\frac{n_{R}^{2}}{\sin^{2}\left(  \Psi_{c}\right)  }\\
	0,
	\end{array}
	\right.  \left.
	\begin{array}
	[c]{c}%
	0\leq\psi_{k,n}\leq\Psi_{c},\\
	\psi_{k,n}>\Psi_{c},%
	\end{array}
	\right.
	\end{equation}
	with $n_{R}\ge0$ being the internal refractive index. Given (${x}_k,{y}_k,{z}_k$) and (${x}_n,{y}_n, {z}_n$) as the coordinates of the $k$-th user and $n$-th LED, respectively, the distance between them can be modeled as $d_{k,n}=\sqrt{(x_{n}-{x}_{k})^{2}+(y_{n}-{y}_{k})^{2}+(z_{n}-{z}_{k})^{2}}$.
\subsection{JD Control}
Regarding the LED array with multiple LEDs and the incorporation of power transfer capability, its energy consumption becomes considerable. In this paper we invoke an efficient JD scheme to control the energy consumption \cite{six}. To this purpose, let $\mathbf{A}$ denote a binary LED selection matrix, where $\mathbf{A} =
\text{diag}(\mathbf{a}) \in \lbrace 0, 1 \rbrace^{N \times N},$ and $\mathbf{a} = [a_{1}, \dots,
a_{N}]^T \in \lbrace 0, 1 \rbrace^{N \times 1}$, i.e., 
\begin{equation}
a_n=\left\{
\begin{array}
[c]{c}%
1,\\
0,
\end{array}
\right. \left.
\begin{array}
[c]{c}%
\text{if the LED $n$ is active},\\
\text{otherwise}.%
\end{array}
\right.
\end{equation}
By this definition, we can declare the number of active LEDs as ${N_{\textrm{a}}} =
\sum\limits_{n = 1}^{{N}} {{a_n}}$.

The JD control scheme includes both AD and SD at the same time, such that the
number of glared LEDs in SD, as well as the uniform DC-bias level of AD are jointly optimized. To
achieve this, a predetermined target dimming level $\eta$ is considered, based on which we can
round-off the number of glared LEDs as follows \cite{seven}
\begin{align} \label{working}
N_a=\big[\eta N\big].
\end{align}
Accordingly, the uniform DC-bias level ${i_{\textrm{DC}}}$ is given by \cite{eight}
\begin{align} \label{AD}
{i_{\textrm{DC}}}=\dfrac{\eta N(I_0-I_l)}{N_a}+I_l,
\end{align}
wherein $I_0=\dfrac{I_l+I_h}{2}$ specifies the original DC-bias that corresponds to AD with all-glared
LEDs and a dimming level of $\eta= 100\%$. Although $N_{\textrm{a}}$ determines the number of glared LEDs, it does not specify the index of
active LEDs at the LED array. Hence, it is required to formulate a network-wide resource allocation
problem, so as to optimize the binary LED selection matrix $\textbf{A}$.
\subsection{Data Rate}
At the receiver side, all users first decode the common received lightweight data stream by considering all private streams as noise. Then, the private received lightweight data stream will be decoded by treating other private streams as noise \cite{Bruno3}. On this basis, the common and private lightweight received data rate for the $k$-th user can be expressed as
   
   \small
	\begin{align}
	 R^{(\textrm{c})}_{k}=\log_{2}\bigg(1+\dfrac{|\bold{h}^{H}_{k}\bold{A}{{\bold{w}}}^{\textrm{(c)}}|^{2}}{\sum_{j=1}^{K}|\bold{h}^{H}_{k}\bold{A}{\bold{w}}_{j}^{\textrm{(p)}}|^{2}+\sigma^{2}_{k}}\bigg), \ \forall k \in \mathcal{K},
	\end{align}
	\normalsize
	and
	\small
	\begin{align}
		R^{(\textrm{p})}_{k}=\log_{2}\bigg(1+\dfrac{|\bold{h}^{H}_{k}\bold{A}{\bold{w}}_{k}^{\textrm{(p)}}|^{2}}{\sum_{j=1,j\neq k }^{K}|\bold{h}^{H}_{k}\bold{A}{\bold{w}}_{j}^{\textrm{(p)}}|^{2}+\sigma^{2}_{k}}\bigg), \ \forall k \in \mathcal{K},
	\end{align}
	\normalsize
	respectively, where ${\bold{h}_k} = {\left[ {{h_{k,1}},{h_{k,2}},...,{h_{k,{N}}}} \right]^T}$ denotes the channel gain vector of user $k$. To ensure that the common stream is successfully decoded, the data rates for common data should satisfy a rate adaptation ${r}^{*}_{k}$, such that
	\begin{align}
		\min_{k} {R^{(\textrm{c})}_{k}}\ge \sum_{k=1}^{K} {r_{k}^{*}}, \ \forall k \in \mathcal{K}.
	\end{align}
	Then, the aggregate system lightweight received data rate can be expressed as
	\begin{align}
		{R}^{\textrm{Agg}}(\mathbf{w}^{\textrm{(c)}}, \lbrace \mathbf{w}_{k}^{\textrm{(p)}} \rbrace_{k \in \mathcal{K}},\mathbf{A},\mathbf{r}^{*} )=\sum_{j=1}^{K}(r_{k}^{*}+{R^{(\textrm{p})}_{k}}),
	\end{align}
where $\mathbf{r}^{*}=[r^{*}_1,r^{*}_2,...,r^{*}_K]^T \in \mathbb{R}^{K\times{1}}$.
\subsection{Energy Harvesting}
The coexistence of AC and DC signals in the photo-current of LEDs, enable users to harvest energy from the DC component, which is blocked by a capacitor. 
The total harvested energy at the $k$-th user from the DC signal of all active LEDs can be expressed as \cite{Guo}
\begin{align}
{P_{k}^\text{Har}} = \sum_{n=1}^{N}\tau {a_n}{V_t}h_{k,n}{i_{\text{DC}}}\ln \left({1 +\frac{\sum_{n=1}^{N}{h_{k,n}{i_{\text{DC}}}}}{{{I_s}}}} \right),
\end{align}
where $V_t$, $\tau$, and $I_s$ are the thermal voltage, the fill factor, and the dark
saturation, respectively. 
Accordingly, the total optical power consumption of the system can be computed as \cite{Tennakoon}
\begin{align}
P^{\text{tot}}(\mathbf{w}^{\textrm{(c)}}, \lbrace \mathbf{w}_{k}^{\textrm{(p)}} \rbrace_{k \in \mathcal{K}},\mathbf{A})=&\zeta\sum_{n=1}^{N}a_n\bigg({w_{n}^{\textrm{(c)}}}+\sum_{k=1}^{K}{w_{k,{n}}^{\textrm{(p)}}}\bigg)\nonumber\\&+P^\text{DC}-\sum\limits_{k = 1}^K{P_k^\text{Har}},
\end{align}
where $P^\text{DC}=\varphi {N_a}{i_{\text{DC}}}$. Moreover, $\zeta\ge 1$ specifies the power for the amplifier efficiency factor, whereas $\varphi$ denotes the conversion factor.	
\subsection{Problem Formulation}
Compared to \cite{three} and \cite{four}, this paper explores the potentials of RSMA and JD control scheme to increase the data rate and control the power consumption, respectively. Particularly, we optimize the aggregate system lightweight received data rate, the beamforming design, LED
selection and RSMA rate adaptation, such that the power budget of the LED array, as well as the QoS and EH thresholds are preserved for all users. Mathematically, the abovementioned network-wide optimization problem is formulated as follows
   			\begin{align}	\label{QoE_max_prob.1}
   				\bold{P}_{1}:&\max_{\mathbf{w}^{\textrm{(c)}}, \lbrace \mathbf{w}^{\textrm{(p)}}_{k} \rbrace_{k \in \mathcal{K}}, \mathbf{A}, \mathbf{r}^{*}}	{R}^{\textrm{Agg}}(\mathbf{w}^{\textrm{(c)}}, \lbrace \mathbf{w}_{k}^{\textrm{(p)}} \rbrace_{k \in \mathcal{K}},\mathbf{A},\mathbf{r}^{*})\nonumber\\&\text{s.t.}~~~ \!~\text{C}_1: \quad \min_{k} {R^{(\textrm{c})}_{k}}\ge \sum_{k=1}^{K} {r_{k}^{*}},~\forall k\in \mathcal{K},\nonumber\\&~~~~~~~\text{C}_2: \quad r_{k}^{*}+{R^{(\textrm{p})}_{k}}\ge{\textrm{QoS}},~\forall k\in \mathcal{K},
   		\nonumber\\&~~~~~~~\text{C}_3: \quad P^\text{tot}\leq P_{\textrm{max}},
   	\nonumber\\&~~~~~~~\text{C}_4: \quad P_k^\text{Har}\ge P^{\textrm{Har}}_\textrm{min},~\forall k\in \mathcal{K},
   	\nonumber\\&~~~~~~~\text{C}_5: \quad a_{n} \in{\{0,1\}},~~\forall n \in \mathcal{N},
   				\nonumber\\&~~~~~~~\text{C}_6:~~~\eta=\dfrac{{N_a}(i_{\textrm{DC}}-I_l)}{N(I_0-I_l)}\times100\%,
   			\nonumber\\&~~~~~~~\text{C}_7:~~~\!\left| {{w_{n}^{\textrm{(c)}}}} \right| + \sum\limits_{k = 1}^K {\left|
{{w_{k,n}^{\textrm{(p)}}}} \right|} \le \Xi, ~\forall n \in
\mathcal{N},
   			\end{align}
   		where $P_{\textrm{max}}$, $P^{\textrm{Har}}_\textrm{min}$, and $\textrm{QoS}$ represent the power budget of the LED array, the minimum homogeneous EH requirement and the minimum homogeneous QoS for all users, respectively. More specifically, $\text{C}_1$ assures the successful signal decoding at all users; $\text{C}_2$ satisfies the QoS for all users; $\text{C}_3$ respects the power budget of the LED array; $\text{C}_4$ specifies the minimum EH requirement for all users; $\text{C}_5$ confines each LED to be either active or inactive; $\text{C}_6$ defines a target required dimming level $\eta$ to be satisfied; Finally, $\text{C}_7$ respects the dynamic range of all LEDs. 
     \\ Concerning the complex domain of $\mathbf{w}^{\textrm{(c)}}$ and $\lbrace \mathbf{w}^{\textrm{(p)}}_{k} \rbrace_{k \in \mathcal{K}}$, the binary domain of $\mathbf{A}$, and the continuous domain of $\mathbf{r}^{*}$, one can clearly claim that this problem is non-convex in the form of mixed integer non-linear programming (MINLP) and belongs to the class of non-deterministic polynomial (NP)-hard problems. The straightforward brute-force method, i.e., the exhaustive search for attaining its globally optimal solution is implausible, considering the coupling and scalability of the problem. Moreover, the classical convex optimization-based solutions mostly rely on time consuming and computationally expensive convex transformations, whereas the wireless environment is quietly dynamic and real-time resource allocation mechanisms are preferred. Instead, we propose a real-time dynamic solution to this problem based on reinforcement learning.
\section{Proposed DRL-based Approach}
In this section, the non-convex problem $\bold{P}_{1}$ with both discrete and continuous variables is firstly reformulated into a model-free MDP, and then a DRL algorithm based on the PPO framework is designed to solve the problem $\bold{P}_{1}$ \cite{PPO1, PPO2, PPO3}. 
\subsection{MDP formulation}  
A $4$-tuple $(\bold{s}_t,\bold{a}_t,r(\mathbf{s}_{t}, \mathbf{a}_{t}),\bold{s}_{t+1})$ is constructed by the MDP formulation, where the current state, the action, the reward function, and the next state are denoted by $\bold{s}_t$, $\bold{a}_t$, $r_{t}$, and $\bold{s}_{t+1}$, respectively. The PPO approach enables the agent to interact with the environment (i.e., the OWC-assisted network), to observe the current state $\bold{s}_t$ from the state space $\mathcal{S}$ and to select the action $\bold{a}_t$ from the action space $ \mathcal{A}$ according to the specific policy with the ultimate aim of maximizing the clipping surrogate objective function ${\mathcal{L}^{\text{CLIP}}}(\cdot)$ that is defined latter. Moreover, based on the formulation of problem $\bold{P}_{1}$, the state, the action, and the reward function are elaborated in the following.  
\subsubsection{State}
The current state $\mathbf{s}_t \in \mathcal{S}$ at time step $t$ constitutes of the main environmental information related to problem $\bold{P}_{1}$ in such a way that allows the policy to enhance and to adapt itself to the dynamic environment. More specifically, the state $\mathbf{s}_{t}$ of the considered system is the set of the common and private rates, the harvested energy, the common and private beamforming vectors as follows   
\begin{align}
&\mathbf{s}_{t} = \\ \nonumber &\Big \lbrace \lbrace R_{k}^{(c)} \rbrace_{k \in \mathcal{K}},   \lbrace R_{k}^{(p)} \rbrace_{k \in \mathcal{K}},   \lbrace P_{k}^{\text{Har}} \rbrace_{k \in \mathcal{K}}, \mathbf{w}^{(c)}, \lbrace \mathbf{w}_{k}^{(p)} \rbrace_{k \in \mathcal{K}} \Big \rbrace.
\end{align}
\subsubsection{Action}
In the proposed PPO approach, action $\mathbf{a}_t \in \mathcal{A}$ at time step $t$ refers to the decisions that an agent takes via an interaction with the considered environment. Furthermore, the action at time step $t$ in problem $\bold{P}_{1}$ consists of both discrete and continuous variables.
\begin{align}
\mathbf{a}_{t} = \Big \lbrace \mathbf{w}^{(c)}, \lbrace \mathbf{w}_{k}^{(p)} \rbrace_{k \in \mathcal{K}}, \mathbf{A}, \mathbf{r}^{*} \Big \rbrace.
\end{align}
\subsubsection{Reward Function}
In particular, the PPO approach is a reinforcement learning method which trains the agents to take suitable decisions in order to maximize the defined clipping
surrogate objective function ${\mathcal{L}^{\text{CLIP}}}(\cdot)$, which contains the reward function $r(\mathbf{s}_{t}, \mathbf{a}_{t})$.
In the optimization problem $\bold{P}_{1}$, the reward function takes both the objective function, i.e., $	{R}^{\textrm{Agg}}(\mathbf{w}^{(c)}, \lbrace \mathbf{w}_{k}^{(p)} \rbrace_{k \in \mathcal{K}},\mathbf{A},\mathbf{r}^{*})$, as well as the constraints of $\bold{P}_{1}$ into account which can be expressed as
\begin{align}
r(\mathbf{s}_{t}, \mathbf{a}_{t}) = {R}^{\textrm{Agg}}(\mathbf{w}^{(c)}, \lbrace \mathbf{w}_{k}^{(p)} \rbrace_{k \in \mathcal{K}},\mathbf{A},\mathbf{r}^{*}) + \sum\limits_{j = 1}^7 {{l_{C_j}}},
\end{align}	
where ${l_{C_j}=\chi_{j} {R}^{\textrm{Agg}}(\mathbf{w}^{\textrm{(c)}}, \lbrace \mathbf{w}_{k}^{\textrm{(p)}} \rbrace_{k \in \mathcal{K}},\mathbf{A},\mathbf{r}^{*}}),$ and the index $j$ corresponds to all constraints, i.e., $\forall j \in \left\{ {1,2,...,7} \right\}$. Besides, $\chi_{j} = 1$, if the ${C_j}$-th constraint is  satisfied and $\chi_{j} = 0$, otherwise. 
\subsection{The PPO-Based Analysis}
In this paper, the PPO algorithm is applied to select actions from both discrete and continuous action spaces, thus solving the non-convex problem \eqref{QoE_max_prob.1}. In particular, 
the PPO is an actor-critic on-policy gradient method which is used to simplify the complex calculation of earlier policy gradient methods, e.g., trust region policy optimization (TRPO). The detailed process of the proposed PPO-based approach is explained as follows. 
\\The main goal in the reinforcement learning is to maximize the expected cumulative reward by considering a  long-term process. Therefore, the cumulative reward at time step $t$ is denoted as 
${R_t} = \sum_{t = 0}^{\infty} \lambda^{t} r(\mathbf{s}_{t}, \mathbf{a}_{t})$, where $\lambda  \in \left[ {0,1} \right)$ represents the discount factor. More specifically, both actor and critic networks are applied to represent the parameterized stochastic policy of action selection denoted as ${\pi _{\thetab}}(\mathbf{a}_{t}|\mathbf{s}_{t})$ and the state-value function ${V_{\phib}(\mathbf{s}_{t})}$, respectively, where $\thetab$ and $\phib$ represent the parameters of the actor and critic networks, respectively. Then, a surrogate objective function based on PPO approach can be expressed as follows 
\begin{align}\label{ggg}
\mathcal{L}\left( \thetab, \mathbf{s}_{t}, \mathbf{a}_{t} \right) = \mathbb{E}\left[\beta_{t}(\thetab) \Omega(\mathbf{s}_{t}, \mathbf{a}_{t}) \right],
\end{align}
where the probability ratio of the current policy and the old one is represented by $\beta_{t}(\thetab) = \pi_{\thetab}(\mathbf{a}_{t}|\mathbf{s}_{t}) / \pi_{\thetab^{\text{old}}}(\mathbf{a}_{t}|\mathbf{s}_{t})$, while  $\thetab^{\text{old}}$ denotes the parameter for the old policy in the actor network. Moreover, the advantage function is given by 
\begin{align}\label{fg}
\Omega(\mathbf{s}_{t}, \mathbf{a}_{t}) = r(\mathbf{s}_{t}, \mathbf{a}_{t}) + \lambda V_{\phib^{\text{old}}}({\mathbf{s}_{t + 1}}) - V_{\phib^{\text{old}}}({\mathbf{s}_{t}}), 
\end{align}
where $\phib^{\text{old}}$ represents the critic network parameter for the old state-value estimation function. To ensure that the updated ${\pi _{\thetab}}(\mathbf{a}_{t}|\mathbf{s}_{t})$ satisfies the trust region constraint, a clipping surrogate objective function can be expressed as 
\begin{align}
&{\mathcal{L}^{\text{CLIP}}}\left(\thetab, \mathbf{s}_{t}, \mathbf{a}_{t} \right) =  \\
&\nonumber \mathbb{E}\Big[\min\left\{
\beta_{t}(\thetab) \Omega ({\mathbf{s}_{t}},{\mathbf{a}_{t}}),
\text{clip}\big(\beta_{t}(\thetab), 1 - \varepsilon ,1 + \varepsilon \big)\Omega ({{\mathbf{s}_{t}}},{\mathbf{a}_{t}})\right\} \Big],
\end{align}
where the clip function is denoted by $\text{clip}(\cdot, \cdot, \cdot)$, while $\varepsilon$ represents a hyper-parameter to restraint $\beta_{t}(\thetab)$ to lie in $\left[ {1 - \varepsilon ,1 + \varepsilon } \right]$. More specifically, the clipping surrogate objective function is iteratively maximized in the proposed PPO approach instead of \eqref{ggg}. Then, a mini-batch stochastic gradient decent (SGD) method updates the corresponding $\thetab$ over $Q$ transitions denoted as $(\bold{s}^q_t,\bold{a}^q_t,r(\mathbf{s}^q_{t},\mathbf{a}^q_{t}),\bold{s}^q_{t+1})$ sampled from an experience pool, which is given by
\begin{align}\label{abc}
\thetab = \thetab^{\text{old}} - {\delta _A}\frac{1}{Q}\sum\limits_{q = 1}^Q {\nabla_{\thetab} \tilde {\mathcal{L}}_q^{\text{CLIP}}} \left( \thetab, \mathbf{s}_{t}^{q}, \mathbf{a}_{t}^{q} \right), 
\end{align} 
where ${\delta _A}$ is the learning rate and $\tilde {\mathcal{L}}_q^{\text{CLIP}}\left( \thetab, \mathbf{s}_{t}^{q}, \mathbf{a}_{t}^{q} \right)$ is the realization of ${\mathcal{L}}^{\text{CLIP}}\left( \thetab, \mathbf{s}_{t}, \mathbf{a}_{t} \right)$ with the $q$-th transition, respectively. 
The mini-batch SGD for updating $\phib$ uses the MSE loss function between $V_{\phib}(\mathbf{s}_{t})$ and $\hat{R}(\mathbf{s}_{t}, \mathbf{a}_{t})$ as follows
\begin{equation}\label{critic_update}
\phib = \phib^{\text{old}} - \delta_C \frac{1}{Q} \sum_{q=1}^{Q} \nabla_{\phib} \big(  V_{\phib}(\mathbf{s}_{t}^{q}) - \hat{R}(\mathbf{s}_{t}^{q}, \mathbf{a}_{t}^{q}) \big)^{2},
\end{equation}
where the learning rate is represented by ${\delta _C}$. Moreover, the target state-value function denoted by $\hat{R}(\mathbf{s}_{t}, \mathbf{a}_{t})$ is given by
\begin{align}\label{target}
\hat{R}(\mathbf{s}_{t}, \mathbf{a}_{t}) = r(\mathbf{s}_{t}, \mathbf{a}_{t}) + \lambda V_{\phib^{\text{old}}}(\mathbf{s}_{t+1}).
\end{align}
\begin{algorithm}[b!]
	\caption{The Proposed PPO-Based Algorithm}
	\label{algorithm1}
	\textbf{Input}: $\Big \lbrace \lbrace R_{k}^{(c)} \rbrace_{k \in \mathcal{K}},   \lbrace R_{k}^{(p)} \rbrace_{k \in \mathcal{K}},   \lbrace P_{k}^{\text{Har}} \rbrace_{k \in \mathcal{K}}, \mathbf{w}^{(c)}, \lbrace \mathbf{w}_{k}^{(p)} \rbrace_{k \in \mathcal{K}} \Big \rbrace$,\\
	\textbf{Output}: $\Big \lbrace \mathbf{w}^{(c)}, \lbrace \mathbf{w}_{k}^{(p)} \rbrace_{k \in \mathcal{K}}, \mathbf{A}, \mathbf{r}^{*} \Big \rbrace$, \\
	\textbf{Initialization}: Initialize the maximum episode $E$ and time step $T$ as well as the parameters of actor and critic networks, i.e., $\thetab$, $\phib$, $\varepsilon$, ${\delta _A}$, and ${\delta _C}$. \\
	Set $\thetab^{\text{old}}=\thetab$ and $\phib^{\text{old}}=\phib$,\\
	\For{episode=1 \KwTo $E$}{
		Initialize state $\mathbf{s}_{t}$,
		\\
		\For{$\text{time step}=1$ \KwTo $T$}{
			Generate action $\mathbf{a}_{t}$ according to $\pi_{\thetab}(\mathbf{a}_{t}|\mathbf{s}_{t})$ in state $\mathbf{s}_{t}$, obtain reward $r(\mathbf{s}_{t}, \mathbf{a}_{t})$ and then observe the new state $\mathbf{s}_{t+1}$, \\
			Store $(\bold{s}_t,\bold{a}_t,r(\mathbf{s}_{t}, \mathbf{a}_{t}),\bold{s}_{t+1})$ in the experience pool,\\
			Calculate the advantage function $\Omega(\mathbf{s}_{t}, \mathbf{a}_{t})$ in \eqref{fg}\\
			\For{$q=1,2,3,...,Q$}{
				Calculate ${\nabla_{\thetab} \tilde {\mathcal{L}}_q^{\text{CLIP}}} \left( \thetab, \mathbf{s}_{t}^{q}, \mathbf{a}_{t}^{q} \right)$ in \eqref{abc},\\
				Calculate $\nabla_{\phib} \big(  V_{\phib}(\mathbf{s}_{t}^{q}) - \hat{R}(\mathbf{s}_{t}^{q}, \mathbf{a}_{t}^{q}) \big)^{2}$ in \eqref{critic_update},\\
				Calculate $\hat{R}(\mathbf{s}_{t}, \mathbf{a}_{t})$ in \eqref{target},}
			Update $\thetab$ and $\phib$ in  \eqref{abc} and \eqref{critic_update}, respectively.
			Update $\thetab^{\text{old}}=\thetab$ and $\phib^{\text{old}}=\phib$.  }}
\end{algorithm}
\\The details of the proposed PPO-based approach are summarized in Algorithm \eqref{algorithm1}. More specifically, the action ${\mathbf{a}_t}$ is generated based on the specific policy in the current state $\mathbf{s}_{t}$ in which the reward $r(\mathbf{s}_{t}, \mathbf{a}_{t})$ is obtained. Furthermore, the transition $(\bold{s}_t,\bold{a}_t,r(\mathbf{s}_{t}, \mathbf{a}_{t}),\bold{s}_{t+1})$ is stored in the experience pool such that Q number of transitions from the experience pool are sampled. In the next step, the advantage function $\Omega (\mathbf{s}_{t}, \mathbf{a}_{t})$ in \eqref{fg} is computed. Finally, the corresponding actor and critic parameters are updated by employing mini-batch SGD. It is worth pointing out that the clipping surrogate objective function in the proposed PPO-based approach ensures that the updated policy satisfies the trust region constraint, thus avoiding the performance collapse.\\
\begin{figure}
	\centering
	\includegraphics[width=9.5cm, height=7cm]{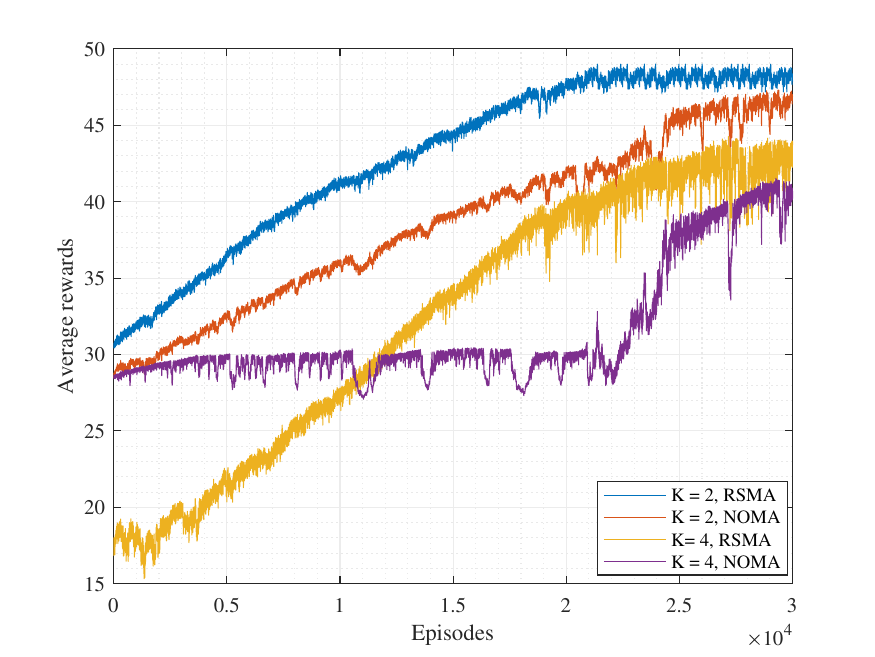}
	\caption{\footnotesize{Convergence: average reward vs the episode number.}}
	\label{MP4}
\end{figure}  
\begin{figure}
	\centering
	\includegraphics[width=9.5cm, height=7cm]{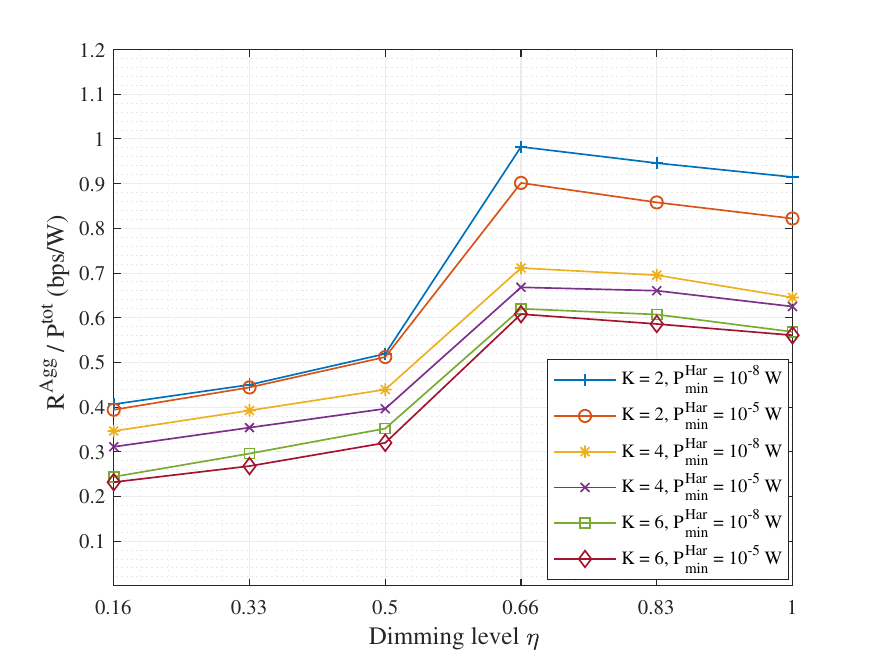}
	\caption{\footnotesize{System EE versus dimming level under $P_{\text{max}}=20~\text{Watts}$ and $\text{QoS}=3~\text{bits/sec}$.}}
	\label{MP1}
\end{figure}
\section{Simulation Results}
In this section, simulation results are presented to assess the performance of the discussed system, within an indoor room of size $8 \times 8 \times 3~m^3$. The random distribution of users should ensure that the coordinates of each user are within the room's dimensions, specifically $0 \le x_k \le 8$, $0 \le y_k \le 8$, and $0 \le z_k \le 1$. In this system, $N=6$ number of LEDs are uniformly distributed on a LED array plane, in which the distance between any two adjacent LEDs sharing the same y-coordinate is set to $2~\text{m}$, while the distance is set to $4~\text{m}$ between any two adjacent LEDs with the same x-coordinate. The simulation parameters are summarized in Table \ref{tab:my_label}, unless otherwise stated.
\\ Fig. 2 displays the convergence behaviour of the proposed solution for both RSMA and NOMA schemes. Notably, more average reward is observed for lower number of users, mainly due to lower imposed inter-user interference. This figure also illustrates that RSMA outperforms NOMA for various number of users at the convergence point, due to more efficient superposition and decoding methodology \cite{PPO2}. 
\\Fig. 3 plots the system energy efficiency (EE) versus various dimming levels. The system EE can be defined as ${R}^{\textrm{Agg}}(\mathbf{w}^{\textrm{(c)}}, \lbrace \mathbf{w}_{k}^{\textrm{(p)}} \rbrace_{k \in \mathcal{K}},\mathbf{A},\mathbf{r}^{*} )/P^{\textrm{tot}}(\mathbf{w}^{\textrm{(c)}}, \lbrace \mathbf{w}_{k}^{\textrm{(p)}} \rbrace_{k \in \mathcal{K}},\mathbf{A})$. Through this figure, we evaluate the system performance for different number of users as well as varying the minimum EH requirement for each user. It is evident that the system EE is maximized around the dimming level of 0.66, representing the optimal trade-off between the data rate and energy consumption of the system. Additionally, for the same number of users, a lower system EE is observed when the minimum EH requirement is higher. For instance, for $K=2$, the baseline related to $P^{\textrm{Har}}_{\textrm{min}}=10^{-5}~\text{Watts}$ achieves lower system EE, compared to the baseline corresponding to $P^{\textrm{Har}}_{\textrm{min}}=10^{-8}~\text{Watts}$. This is due to the more stringent constraint $\textrm{C}_4$, leading to a more limited feasible set for problem $\bold{P}_1$. 
\\Additionally, Fig. 4 evaluates the average system data rate for various minimum QoS for users. The baselines constitute either RSMA or NOMA, with various dimming levels. We can observe that our proposed scheme with RSMA outperforms the baseline with NOMA. It is evident that increasing the minimum QoS leads to a reduction in the average system data rate for both RSMA and NOMA baseline schemes, under the same justification stated for Fig. 3.
\\Furthermore, Fig. 5 illustrates the average system data rate versus the minimum EH requirement, where baselines constitute either RSMA and NOMA with different dimming levels. As can be seen, a higher dimming level achieves more average system data rate, regardless of multiple access scheme (i.e., RSMA or NOMA). Thus, RSMA consistently outperforms NOMA in achieving a higher average system data rate for a specific minimum EH requirement. 
\begin{figure}
	\centering
	\includegraphics[width=9.5cm, height=7cm]{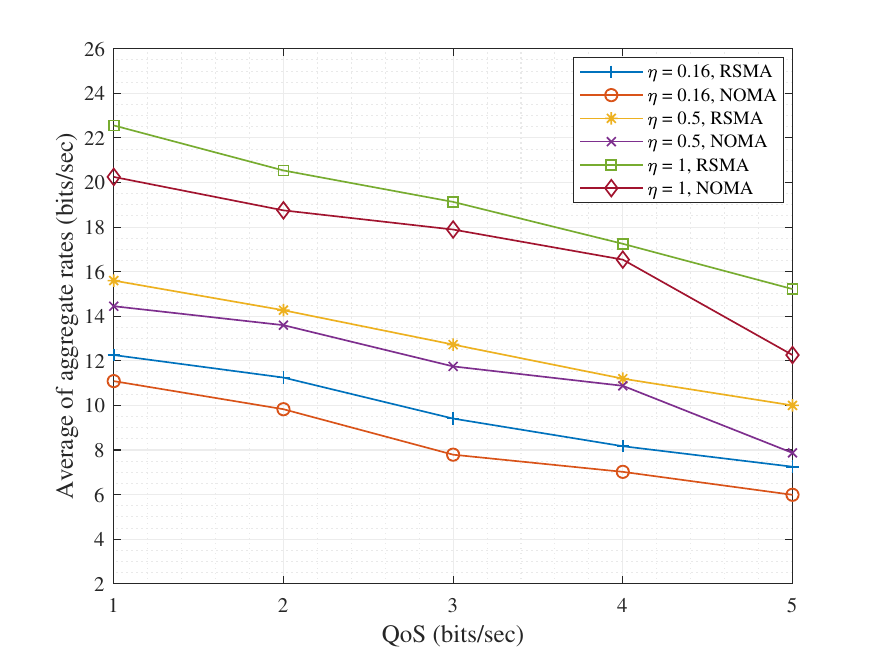}
	\caption{\footnotesize{Average system data rate versus the minimum QoS of users, under $P_{\text{max}}=20~\text{Watts}$, $K=4$, and $P^\text{Har}_{\text{min}}=10^{-8}~\text{Watts}$.}}
	\label{MP3}
\end{figure} 
\begin{figure}
	\centering
	\includegraphics[width=9.5cm, height=7cm]{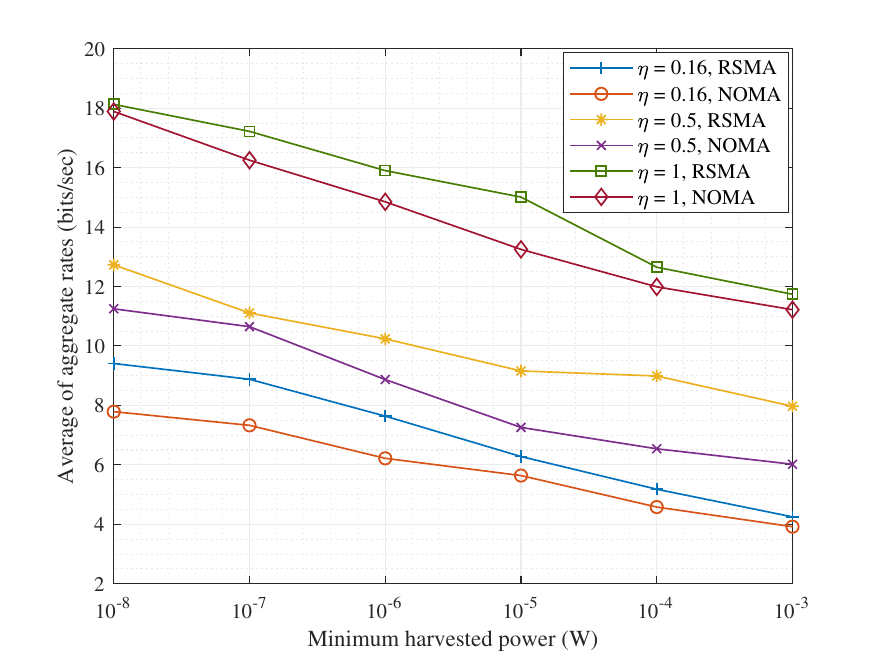}
	\caption{\footnotesize{Average system data rate versus the minimum EH requirement under $P_{\text{max}}=20~\text{Watts}$, $K=4$, and $\text{QoS}=3~\text{bits/sec}$.}}
	\label{MP2}
\end{figure} 
\begin{table}[t!] 
	\vspace{-1 em}
	\centering
	\caption{Simulation Parameters}
	\label{tab:my_label}
	\vspace*{5pt}
	\begin{tabular}{ l|l||l|l }
		\textbf{Parameter} & \textbf{Value} & \textbf{Parameter} & \textbf{Value} \\
		\hline
		\hline		
		\hline
		$\Psi_{c}$   & $~60^{\circ}$ &
		$n_{R}$ & $1.5$\\ 
		\hline $ \phi_{1/2}$ & $~60^{\circ}$ &
		$\text{A}^\text{OWC}$& $1~\text{cm}^2$ \\ 
		\hline $I_h$ & $10~\text{mA}$ &
		$I_l$& $0~\text{A}$ \\
		\hline $I_0$ & $5~\text{mA}$  &
		$I_s$ & $10^{-9}~\text{A}$  \\
		\hline $V_t$ & $25~\text{mA}$ &
		$\zeta$ & $1.2$ \\
		\hline $\varphi$ & $1$ &
		$\tau$ & $0.75$ \\
		\hline $N$ & $6$  &
        $P_\text{max}$ & $20~\text{W}$ \\
		\hline
	\end{tabular}
	\label{tab2}
	\vspace{-0.75 em}
\end{table}
\section{Conclusion}
In this paper, the performance of a multi-LED OWC system with SLIPT technology is investigated, where a JD control scheme is proposed to reduce its energy consumption, and RSMA technology is explored to increase its data rate. We formulated a resource allocation problem and proposed a dynamic real-time solution based on reinforcement learning. We numerically found the optimal dimming level in this system to achieve a trade-off between the data rate and energy consumption.
\section{Acknowledgment} 
The work of P. D. Diamantoulakis was supported by the Hellenic Foundation for Research and Innovation (H.F.R.I.) under the “3rd Call for H.F.R.I. Research Projects to support Post-Doctoral Researchers” (Project Number: 7280).
The work of G. K. Karagiannidis was implemented in the framework of H.F.R.I call “Basic research Financing (Horizontal support of all Sciences)” under the National Recovery and Resilience Plan “Greece 2.0” funded by the European Union – NextGenerationEU (H.F.R.I. Project Number: 15642).

\end{document}